\documentclass[12pt,a4paper]{article}
\pdfoutput=1

\usepackage{a4wide}

\usepackage{latexsym}
\usepackage{epsf}
\usepackage{amssymb}
\usepackage{graphicx}
\usepackage{amsmath, cite}
\usepackage{amsmath,amssymb,amsthm}
\usepackage{verbatim}
\usepackage{hyperref}
\usepackage{amsmath}
\usepackage{amsfonts}
\usepackage{xfrac}
\usepackage{slashed}
\usepackage{tikz}
\usepackage{cancel}
\usepackage{bbm}

\usepackage[utf8]{inputenc}

\usepackage{fancyhdr}
\usepackage{datetime}

\usepackage{makeidx}

\newcommand{\Tr}{\textrm{Tr}\,}
\newcommand{\Trs}[1]{\textrm{Tr}_{#1}\,}

\newcommand{\w}{\wedge}

\newcommand\varpm{\mathbin{\vcenter{\hbox{%
  \oalign{\hfil$\scriptstyle+$\hfil\cr
          \noalign{\kern-.3ex}
          $\scriptscriptstyle({-})$\cr}%
}}}}

\newcommand\varmp{\mathbin{\vcenter{\hbox{%
   \oalign{\hfil$\scriptstyle-$\hfil\cr
           \noalign{\kern-.3ex}
          $\scriptscriptstyle({+})$\cr}%
}}}}

\newcommand{\be}{\begin{equation}}
\newcommand{\ee}{\end{equation}}
\newcommand{\bea}{\begin{eqnarray}}
\newcommand{\eea}{\end{eqnarray}}

\usetikzlibrary{arrows,automata,positioning,calc,trees,decorations.pathmorphing,decorations.markings}

\usetikzlibrary{arrows,automata,positioning,calc,trees,decorations.pathmorphing,decorations.markings}

\begin{document}
\numberwithin{equation}{section}
\begin{flushright}
\small
IPhT-T17/027\\
IFT-UAM/CSIC-17-023
\normalsize
\end{flushright}

\vspace{1 cm}

\begin{center}
{\LARGE T-branes and Matrix Models}

\vspace{2.6 cm} {\large Iosif Bena$^{\dagger}$, Johan Bl{\aa}b{\"a}ck$^{\dagger}$, Raffaele Savelli$^*$}\\

\vspace{0.85 cm}{$^{\dagger}$ Institut de Physique Th{\'e}orique, Universit{\'e} Paris Saclay, CEA, CNRS,\\ Orme des Merisiers, F-91191 Gif-sur-Yvette, France\\ $^{*}$ Instituto de F\'isica T\'eorica UAM-CSIC, Cantoblanco, 28049 Madrid, Spain}
\\\vspace{0.4 cm}

\vspace{0.45cm} {\small\upshape\ttfamily iosif.bena, johan.blaback @ cea.fr, raffaele.savelli @ uam.es} \\

\vspace{3cm}


\textbf{Abstract}
\end{center}

\begin{quotation}

We find that the equations describing T-branes with constant worldvolume fields are  identical to the equations found by Banks, Seiberg and Shenker twenty years ago to describe longitudinal five-branes in the BFSS matrix model. Besides giving new ways to construct T-brane solutions, this connection also helps elucidate the physics of T-branes in the regime of parameters where their worldvolume fields are larger than the string scale. We construct explicit solutions to the Banks-Seiberg-Shenker equations and show that the corresponding T-branes admit an alternative description as Abelian branes at angles.
\end{quotation}

\newpage




\section{Introduction}

\begin{figure}[h!]
  \begin{center}
    \includegraphics[scale=1.2]{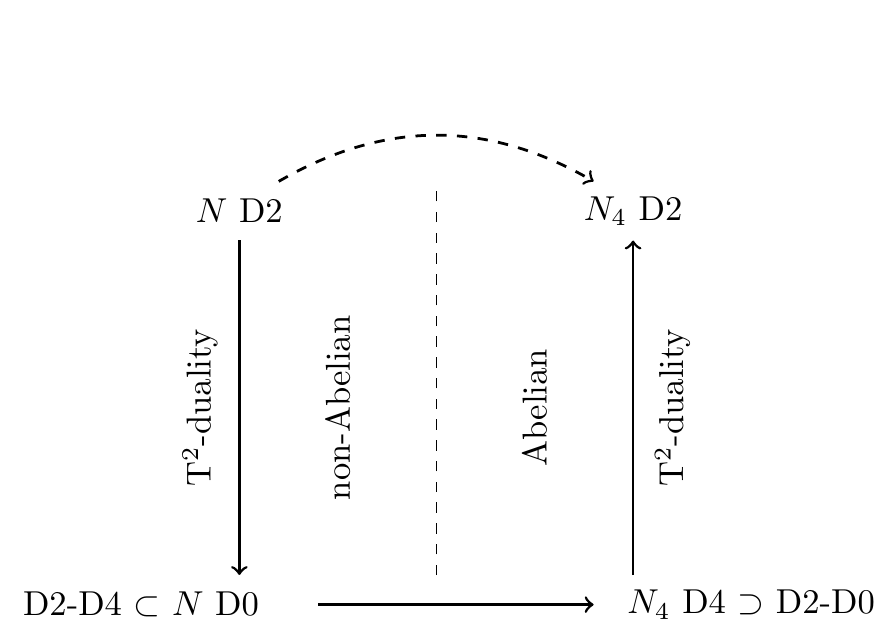}
  \end{center}
  \caption{The summary of our construction. }
\end{figure}\label{MasterFigure}

T-branes are supersymmetric brane configurations in which two scalars and the worldvolume flux acquire non-commuting expectation values. They were first introduced in  \cite{Cecotti:2010bp}, and have since received a fair bit of interest, with reasons ranging from their {fundamental} structure to the attractiveness of their low-energy features for model-building in string phenomenology \cite{Chiou:2011js,Donagi:2011jy,Donagi:2011dv,Marsano:2012bf,Font:2012wq,Font:2013ida,Anderson:2013rka,DelZotto:2014hpa,Collinucci:2014qfa,Collinucci:2014taa,Marchesano:2015dfa,Cicoli:2015ylx,Carta:2015eoh,Collinucci:2016hpz,Ashfaque:2017iog,Anderson:2017rpr}.

Despite this, several aspects of T-branes have remained quite mysterious. In particular, the presence of non-Abelian scalar vevs seems to hint at a possible interpretation of T-branes in terms of higher-dimensional branes, similar in spirit to the Myers effect \cite{Myers:1999ps}.

In \cite{Bena:2016oqr}, the authors and Minasian have shown that for certain classes of T-branes such an interpretation is incorrect: In the regime of large worldvolume fields (in string units) these T-branes appear rather to be described by Abelian branes wrapping certain holomorphic surfaces, whose curvature encodes the original T-brane data. Roughly speaking, the non-Abelian vacuum profiles of T-branes give rise to brane bending, and not to brane polarization.

The original purpose of the present investigation was to understand how universal the connection found in \cite{Bena:2016oqr} is, by investigating other classes of T-brane solutions and, in particular, those with constant worldvolume fields. However, a surprise awaited us:  We discovered that the Hitchin system describing this class of T-branes is exactly the same as the system of equations that was found by Banks, Seiberg and Shenker in \cite{Banks:1996nn} to describe longitudinal five-branes in the BFSS matrix model \cite{Banks:1996vh} (see also \cite{Ganor:1996zk,Berkooz:1996is,Castelino:1997rv}). Upon reduction to type IIA string theory, the Banks-Seiberg-Shenker equations describe a non-Abelian configuration of D$0$-branes that preserves eight supercharges and carries D$2$ and D$4$ charges.

The fact that these equations are identical points to the existence  of a more profound connection, which has to do with the fact that both the BFSS matrix model and the Hitchin system describing T-branes come from reductions of ten-dimensional super-Yang-Mills theory to lower dimensions: The BFSS matrix model is the reduction of this theory to a particular one-dimensional matrix quantum mechanics, while the Hitchin system arises from an intermediate two-dimensional compactification of the self-duality equations of the super-Yang-Mills theory \cite{Hitchin:1986vp}.

Armed with this connection, one can use the extensive technology developed in the good old matrix-model days to construct, rather straightforwardly, several solutions of T-branes with constant fields. As we will show, to obtain such T-branes one has to consider infinite matrices, and we construct a map between these T-branes and their Abelian counterparts following a path similar to that of \cite{Bena:2016oqr}:
The system of equations we obtain in the T-brane frame is mapped to a dual system via two T-dualities along the worldvolume of the T-brane. The resulting dual system describes a particular D$0$-D$2$-D$4$\footnote{We will mostly refer to the T-brane as made of D$2$-branes in this paper, for historical matrix-model reasons. This is however done without loss of generality; all the same conclusions can be drawn for any D$p$-brane stack for $p=2,\ldots,7$.} configuration from the perspective of D0-branes with non-Abelian worldvolume-scalar vacuum expectation values. The same system can be described as two or more D4-branes with Abelian worldvolume fluxes, which, when T-dualized back to the original frame, give rise to several intersecting D2 branes.

In ``black-hole'' language, the map between the D0 and the D4 descriptions that we construct is not a microscopic map, but a macroscopic one. To see this, it is important to recall that the D0-D4 system has a very large number of states, of order $e^{2 \pi \sqrt{2 N_0 N_4}}$, and each of these states can be in principle described either from a D0-brane perspective, as a vacuum configuration where the scalars of the D0-brane worldvolume have non-commutative vacuum expectation values, or from the D4 perspective, as an instanton configuration on the D4-brane worldvolume. The precise map between individual microstates is only known for a few very specific microstates, and requires in general pretty complicated technology. Our purpose is not to construct this detailed microscopic map, but rather to identify ensemble representatives that have the same overall D4, D2, and D0 charges.

The Abelian system that we find is then brought back to the original T-brane frame by reversing the two T-dualities. At the end of this last step, we recover a D$2$-brane system, which gives the Abelian description of the original non-Abelian {T-brane system}. Thus, we find the same underlying physics as in \cite{Bena:2016oqr}: {T-brane configurations of stacks of D$p$-branes can be mapped to Abelian systems of D$p$-branes}.

Our map can clearly be made more precise, both on the lower side of Figure \ref{MasterFigure} (by finding for example relations between three-point functions in the matrix-model description and D0 density modes in the D4 worldvolume description) and on the upper side of Figure \ref{MasterFigure}  (by relating the T-brane data to the precise shape of the holomorphic curves wrapped by D2-branes), and we leave this investigation for future work.

The paper is organized as follows. In Section \ref{sec:set} we present our T-brane system and map it to the Banks-Seiberg-Shenker system in Matrix Theory through two T-dualities. In the language of Figure \ref{MasterFigure} we start in the upper left corner, and move downwards. In the lower left corner we construct an explicit solution, which is presented in Section \ref{sec:sol}. We work out a map between the lower left and right corners in Section \ref{sec:abel}, and present the resulting D$4$-brane solution. In Section \ref{sec:return}, we move to the upper right corner of Figure \ref{MasterFigure}, where we construct the Abelian intersecting-brane configuration that corresponds to our original T-brane. The paper is concluded with some observations in Section \ref{sec:inter}.

\section{From T-branes to Matrix Theory}\label{sec:set}

T-branes preserving eight supercharges are non-trivial solutions of the so-called Hitchin system:
\begin{subequations}\label{HitchinSystem}
\begin{eqnarray}
\bar{\partial}_A\,\Phi&=&0\,, \label{HF}\\
F+[\Phi,\Phi^\dagger]&=&0\,.\label{HD}
\end{eqnarray}
\end{subequations}
This system is defined on $\mathbb{C}_w \times \mathbb{C}_z$, parametrized by the complex coordinates $w$ and $z$, that are parallel and transveral to the D-brane directions, respectively. The anti-holomorphic part of the anti-Hermitian $SU(N)$ gauge connection, $A_{\bar{w}}$, has a field strength $F=\partial A_{\bar{w}}+\bar{\partial}A_w+[A_w,A_{\bar{w}}]$, where $\bar{\partial}_A \equiv  \bar{\partial}+[A_{\bar{w}},\cdot]$. Moreover, $\Phi$, usually called the ``Higgs field'', is the complex combination of two of the worldvolume scalars of the D-brane stack, and is a holomorphic (1,0)-form valued in the adjoint representation of the gauge group.

Before beginning we would like to make some preliminary observations on these equations. T-brane configurations are characterized by a non-trivial commutator $[\Phi,\Phi^\dagger]$ and, because of the cyclicity of the trace, have a traceless worldvolume flux. The field $\Phi$, however, is not necessarily traceless.
In this paper we are interested in T-branes that have constant worldvolume fields, for which the equations above are written solely in terms of commutators:
\begin{subequations}\label{HitchinSystemC}
 \begin{eqnarray}
   \label{HFC} [A_{\bar{w}},\Phi] &=&0\,,\\
   \label{HDC} [A_w, A_{\bar{w}}] + [\Phi, \Phi^\dagger] &=&0 \,.
  \end{eqnarray}
\end{subequations}
Since, as we will reiterate below, these equations can only be non-trivially solved for infinite matrices, all commutators can in principle admit a non-trivial trace. However, since we have finite-$N$ T-branes in mind, we will still keep the commutators appearing in \eqref{HDC} traceless, whereas we will allow for non-trivial traces in \eqref{HFC} (as we will see, these just give rise to additional harmless brane charges, without spoiling supersymmetry).

Upon expressing the complexified fields $A_w$ and $\Phi$ in terms of their Hermitian components\footnote{From now on we only consider the matrix-valued coefficients of the differential forms, but refrain from introducing a new notation.}
\begin{equation}
  \begin{split}
    A_w   &= -\frac{1}{2}(A_3 + i A_4)\,,\\
    A_{\bar{w}}   &= \frac{1}{2}(A_3 - i A_4)\,,\\
    \Phi  &= \frac{1}{2}(\Phi_1 + i \Phi_2)\,,
  \end{split}
\end{equation}
{the system \eqref{HitchinSystemC} becomes}
\begin{equation}\label{HitchinSystemCR}
  \begin{split}
    \left\{\begin{array}{r} [\Phi_1, A_4]\\ {[}\Phi_1, A_3{]} \end{array}\right. &\begin{array}{r}=\ [\Phi_2, A_3]\,, \\ =\ {[}A_4, \Phi_2{]}\,, \end{array}\\
 [\Phi_1, \Phi_2]\ \,&=\,\,\, [A_3, A_4]\,,
  \end{split}
\end{equation}
where the first two equations come from the anti-Hermitian and Hermitian parts of \eqref{HF} respectively, and the last comes from \eqref{HDC}.

\begin{table}
  \begin{center}
    \begin{tabular}{c||c|c|c|c}
                & $\mathbb{R}^{p-2,1}$  & $\mathbb{R}^{7-p}$  &  {$\mathbb{C}_w \to \mathbb{R} \times \mathbb{R}$} &   $\mathbb{C}_z$\\
                \hline
                \hline
      T-brane   & $\times$              &                     & $\times$      &\\
                \hline
                &                       &                     & $A_3$ $A_4$   & $\Phi$\\
                \hline
                \hline
      T-dual    &                       &                     & $\downarrow\ $$\ \downarrow$ & \\
                \hline
                \hline
      dual brane& $\times$              &                     &               & \\
                \hline
                &                       &                     &$\Phi_3$ $\Phi_4$ & $\Phi$
    \end{tabular}
  \end{center}
  \caption{Illustration of the two T-dualities.}
  \label{tab:T-duals}
\end{table}

{Following a train of logic similar to that of \cite{Bena:2016oqr}, we now T-dualize the T-brane equations \eqref{HitchinSystemCR} twice along the worldvolume directions $3$ and $4$ (see Table \ref{tab:T-duals}).} This maps the gauge potentials $A_{3,4}$ into worldvolume scalars $\Phi_{3,4}$, and the T-brane equations become:
\begin{equation}\label{eq:master}
  \begin{split}
    [\Phi_1, \Phi_4] &= [\Phi_2, \Phi_3]\,,\\
    [\Phi_1, \Phi_3] &= [\Phi_4, \Phi_2]\,,\\
    [\Phi_1, \Phi_2] &= [\Phi_3, \Phi_4]\,.
  \end{split}
\end{equation}
or more concisely
\begin{equation}\label{eq:mastersmall}
    \frac{1}{2}\sum_{i,j}\epsilon_{ijkl}[\Phi_i, \Phi_j] = [\Phi_k, \Phi_l].\\
\end{equation}

The first surprise in our investigation is that this system is exactly the same as the Banks-Seiberg-Shenker system of equations  \cite{Banks:1996nn} that describes longitudinal five-branes in the BFSS matrix model \cite{Banks:1996vh}. Upon compactifying to type IIA string theory, these equations describe multiple D$0$-branes dissolved into D$4$-branes (with extra possible D$2$ charges) from the perspective of the worldvolume non-Abelian Born-Infeld action of the D$0$-branes. As noted in \cite{Banks:1996nn}, this system of equations admits no non-trivial solutions in terms of finite matrices, and hence to proceed we will henceforth use infinite matrices. We will further discuss the relevance of this construction for finite-$N$ T-branes in Section \ref{sec:inter}.

To demonstrate that indeed this system contains D$2$-branes as well as D$4$-branes, we can derive an expression for their charge densities from the Wess-Zumino part of the non-Abelian Born-Infeld action of $N$ D0-branes \cite{Myers:1999ps}:
\begin{equation}
  S^{\textrm{D}0}_{\textrm{WZ}} =
  \mu_0 \int C_1
  + \left(- i \frac{\mu_0 \lambda}{L^2} \Tr [\Phi_i,\Phi_j]\right) \int  C_3 ^{ij}
  + \left(- \frac{\mu_0 \lambda^2}{2 L^4} \epsilon^{ijkl} \Tr \Phi_l \Phi_k \Phi_j \Phi_i \right) \int C_5^{1234}\,.
\end{equation}
where  $\lambda = 2\pi \ell_s^2 = 2\pi \alpha'$, $\mu_p = 2\pi/(2\pi \ell_s)^{p+1}$, and the extra factors of $L$ come from the fact that the volume\footnote{We are here quite liberal with the use of the phrase \emph{volume}, as $L$ is derived from the topological Wess-Zumino term: It does not strictly give a volume but rather gives information about the boundaries. However, for the flat branes we are considering here, these two agree and we will keep on slightly abusing the nomenclature.} of the D$2$-branes is $L^2$ and the volume of the D$4$-branes is $L^4$.

The induced numbers of D$p$-branes, $N_p$, are given by the electric couplings between the D0-brane fields and $ C_{p+1}$
\begin{equation}
  S_{\textrm{WZ}}^{\textrm{D}0} = \ldots + \mu_p N_p \int C_{p+1} + \ldots\,,
\end{equation}
and to express them in terms of matrices it is convenient to define the dimensionless quantities $\tilde{\Phi}\equiv\sqrt{\lambda}\Phi$ and $K\equiv L/\sqrt{2\pi\lambda}$. The D2 and D4 numbers are then
\begin{equation}\label{eq:0branenr}
  \begin{split}
    N_2^{ij} &= -i \frac{1}{K^2} \Tr \left[\tilde{\Phi}_i,\tilde{\Phi}_j\right]\,,\\
    N_4 &= - \frac{1}{2 K^4} \epsilon^{ijkl} \Tr \tilde{\Phi}_l \tilde{\Phi}_k \tilde{\Phi}_j \tilde{\Phi}_i\,,
  \end{split}
\end{equation}
where the $ij$ superscript on the number denoting D$2$-branes signify their orientation, according to the left hand side of Table \ref{tab:system}. From now throughout the rest of this paper we will exclusively use the dimensionless fields $\tilde{\Phi}_i$, but proceed to drop the tilde in order to un-clutter the formulae. Note that $K$ can be thought of as the dimensionless size of the box in which our D0-branes are distributed, and, like $N$, must be taken to infinity. Equations \eqref{eq:0branenr} and the cyclicity of the trace make it clear that to be able to induce non-trivial D2 charges one has to use infinite matrices $\Phi_i$. As explained at the beginning of this section, we will consider T-branes for which the $N_2^{12} = N_2^{34} = 0$ because of the necessity of the tracelessness of Equation (\ref{HD}) for finite matrices. We will impose this condition in order not to introduce new features unrelated to T-branes. However, at the same time, we will allow ourselves to ``dress'' the T-brane with the other D$2$-brane charges, $N_2^{13} = N_2^{42}$ and $N_2^{14} = N_2^{23}$, since these correspond to finite traces of various terms in equation \eqref{HF}, which are allowed for finite matrices.

\begin{table}
  \begin{center}
    \begin{tabular}{r|cccc}
           & 1        & 2         & $\tilde{3}$         & $\tilde{4}$ \\
      \hline
      {D$0$} & -        & -         & -         & - \\
      \hline
      D$4$ & $\times$ & $\times$  & $\times$  & $\times$ \\
      \hline
     \textcolor{red}{\underline{D$2$}} & $\times$ & $\times$  & -         & - \\
      \textcolor{red}{\underline{D$2$}} & -        & -         & $\times$  & $\times$ \\
      D$2$ & $\times$ & -         & $\times$  & - \\
      D$2$ & -        & $\times$  & -         & $\times$ \\
      D$2$ & $\times$ & -         & -         & $\times$ \\
      D$2$ & -        & $\times$  & $\times$  & - \\
      \end{tabular}
      $\qquad$
      \begin{tabular}{r|cccc}
             & 1 & 2 & 3 & 4\\
        \hline
        {D$2$} & -        & -         & $\times$  & $\times$ \\
        \hline
        D$2$  & $\times$ & $\times$  & -         & -\\
        \hline
        \textcolor{red}{\underline{D$4$}} & $\times$ & $\times$  & $\times$  & $\times$ \\
        \textcolor{red}{\underline{D$0$}} & -        & -         & -         & - \\
        D$2$ & $\times$ & -         & -         & $\times$ \\
        D$2$ & -        & $\times$  & $\times$  & - \\
        D$2$ & $\times$ & -         & $\times$  & - \\
        D$2$ & -        & $\times$  & -         & $\times$ \\
        \end{tabular}
  \end{center}
  \caption{In the left table we display the branes present in a general solution of (\ref{eq:master}). To the right is the resulting branes after reversing the T-dualities depicted in Table \ref{tab:T-duals}, e.g. the T-brane frame. The branes colored in red and underlined are not present in a T-brane solution.}
  \label{tab:system}
\end{table}

\section{Finding a solution}\label{sec:sol}

The goal of this section is to find an explicit solution to the system (\ref{eq:master}). The building blocks for constructing solutions to this system of equations are two infinite Hermitian traceless matrices $D$ and $X$, analogous to momentum and position operators,  satisfying the relation
\begin{equation}\label{FundComm}
  [D,X] = i \mathbb{I}_M\,,
\end{equation}
where the size of the matrices, $M$, is actually infinity, but we keep track of it for the purpose of making the normalizations clear.  Explicitly, these matrices can be constructed from the creation and annihilation operators of Quantum Mechanics via
\begin{equation}
  D \equiv \frac{1}{\sqrt{2}} \left(a + a^\dagger\right)\,,\quad X \equiv \frac{i}{\sqrt{2}} \left(a^\dagger - a\right)\,,
\end{equation}
with\footnote{This particular choice is not compulsory, there exist other types of infinite matrices that can represent $a$ and $ a^\dagger$, but this choice makes the calculations more straightforward.}
\begin{equation}
  a^\dagger = \begin{pmatrix}0&0&0&\dots &0&\dots \\{\sqrt {1}}&0&0&\dots &0&\dots \\0&{\sqrt {2}}&0&\dots &0&\dots \\0&0&{\sqrt {3}}&\dots &0&\dots \\\vdots &\vdots &\vdots &\ddots &\vdots &\dots \\0&0&0&0&{\sqrt {n}}&\dots &\\\vdots &\vdots &\vdots &\vdots &\vdots &\ddots \end{pmatrix}\,.
\end{equation}

Our dynamics takes place in four dimensions and we can construct four-dimensional momentum and position operators of size $M^4\times M^4 = N\times N$:
\begin{equation}\label{eq:decomp}
  \begin{split}
    D_i &= \bigotimes_{j=1}^4 \left( (1-\delta_{ij}) \mathbb{I}_M + \delta_{ij} D\right)\,,\\
    X_i &= \bigotimes_{j=1}^4 \left( (1-\delta_{ij}) \mathbb{I}_M + \delta_{ij} X\right)\,,\\
  \end{split}
\end{equation}
which {satisfy}
\begin{equation}
  [D_i,X_j] = i \delta_{ij} \times \mathbb{I}_M \otimes \mathbb{I}_M \otimes \mathbb{I}_M \otimes \mathbb{I}_M = i \delta_{ij}\mathbb{I}_N\,,
\end{equation}

We can now construct Ans\"atze for the matrices $\Phi_i$ in terms of $D_i$ and $X_i$. As already mentioned, the goal is to find a solution that has non-vanishing charge for all the D$2$-branes except the D$2_{12}$ and D$2_{34}$, but still have $[\Phi_1,\Phi_2] = [\Phi_3,\Phi_4]$. This can be achieved for example by the following three-parameter family of solutions
\begin{equation}\label{eq:na-sol}
\begin{split}
  \Phi_1 &= D_1 - A_{14} X_4 - A_{13} X_3 - \frac{\gamma}{\sqrt{2M}} (X_2 X_4 + X_1 X_3)\,,\\
  \Phi_2 &= D_2 + A_{13} X_4 - A_{14} X_3\,,\\
  \Phi_3 &= D_3\,,\\
  \Phi_4 &= D_4 + \frac{\gamma}{\sqrt{2M}} (X_3 X_4 - X_1 X_2)\,,
\end{split}
\end{equation}
where $A_{13}, A_{14},\gamma$ are constants whose physical meaning will be clear shortly. The matrices $\Phi_i$ in \eqref{eq:na-sol} have the commutators
\begin{equation}
  \begin{split}
    [\Phi_1, \Phi_2] = [\Phi_3,\Phi_4] &= i\frac{\gamma}{\sqrt{2M}} X_4\,,\\
    [\Phi_1, \Phi_3] = [\Phi_4,\Phi_2] &= iA_{13} \mathbb{I}_N + i\frac{\gamma}{\sqrt{2M}} X_1\,,\\
    [\Phi_1, \Phi_4] = [\Phi_2,\Phi_3] &= iA_{14} \mathbb{I}_N \,,\\
  \end{split}
\end{equation}
and hence equation \eqref{eq:0branenr} implies that the D$2$-brane charges are
\begin{equation}\label{eq:ND2}
  \begin{split}
    N_2^{(12)} &= N_2^{(34)} = 0\\
    N_2^{(13)} &= N_2^{(42)} = A_{13} \frac{N}{K^2}\,,\\
    N_2^{(14)} &= N_2^{(23)} = A_{14} \frac{N}{K^2}\,.
  \end{split}
\end{equation}
These charges do not depend on $\gamma$, because the $X_i$ are traceless. However, the D4-brane charge does depend on $\gamma$:
\begin{equation}\label{eq:ND4}
  N_4 = \frac{N}{K^4} \left( A_{14}^2 + A_{13}^2 + \gamma^2 \right)\,.
\end{equation}
These dependences highlight the crucial role played by the parameter $\gamma$ of our family of solutions. If a solution allows the following decomposition of the trace
\begin{equation}
  \begin{split}\label{eq:BPS1}
    N_4 &= - \frac{1}{2 K^4} \epsilon^{ijkl} \Tr \Phi_l \Phi_k \Phi_j \Phi_i
         = - \sum_{\textrm{D}2\textrm{-pairs}} \frac{1}{K^4}\Tr \{\star[\Phi_i, \Phi_j], [\Phi_i,\Phi_j]\}\\
        &= \frac{1}{N} \sum_{\textrm{D}2\textrm{-pairs}} \left(-i\frac{1}{K^2} \Tr [\Phi_i,\Phi_j]\right) \left(-i\frac{1}{K^2} \Tr [\Phi_i,\Phi_j]\right) \\
        &= \frac{1}{N} \sum_{\textrm{D}2\textrm{-pairs}} \left(N^{ij}_2\right)^2\,,
  \end{split}
\end{equation}
then it features supersymmetry enhancement and preserves $16$ supercharges. As one can see from \eqref{eq:ND4}, this condition is broken by $\gamma$, and therefore only the solutions with non-zero $\gamma$ will preserve just $8$ supercharges. Hence, it is $\gamma$ that gives to our solution a T-brane character, because it is the only parameter appearing in equation \eqref{HD}. On the other hand, the parameters $A_{13}$ and $A_{14}$ are only there to ``dress'' the T-brane with additional D2 charges, without spoiling its features.

Let us now work out the finite physical quantities of our family of solutions. We have a number $N$ of D$0$-branes which we are implicitly sending to infinity. These branes are distributed over an infinite four-dimensional space of volume $K^4$, and the appropriate finite quantity in our solution is the average density of D$0$-branes:
\begin{equation}\label{D0dens}
  \rho_0 = \frac{N}{K^4} < \infty\,.
\end{equation}
{The same can be said for D2-branes: they are distributed in a subspace of volume $K^2$ so their number is infinite but their density is finite:}
\begin{equation}\label{D2dens}
  \rho_2^{ij} = \frac{N^{ij}_2}{K^2} = \rho_0 A_{ij} < \infty\,.
\end{equation}
Since the D4-branes wrap the whole four-dimensional space, the D4 charge $N_4$ is the same as the D4 density, $\rho_4=N_4$, and hence Equation \eqref{eq:ND4} can be rewritten using finite brane densities
\begin{equation}\label{D4dens}
  \rho_4 \rho_0 = \sum_{\textrm{D}2\textrm{-pairs}} (\rho_2^{ij})^2 + \rho_0^2 \gamma^2\,.
\end{equation}

To summarize, we may formulate our non-Abelian picture solely in terms of finite quantities as follows. We start by fixing the quantity $\rho_0$, which is the analogue of the size of finite-dimensional matrices. By rescaling our infinite matrices, we can make it appear in the fundamental commutation relation \eqref{FundComm}, so that
\begin{equation}\label{FundCommResc}
\frac{1}{N}\Tr[D_i, X_j]=i \delta_{ij} \rho_0\,.
\end{equation}
Now, the three-parameter family of explicit solutions is formally given by \eqref{eq:na-sol}, from which, by computing the relevant traces and using \eqref{FundCommResc}, we can extract the D2-brane densities \eqref{D2dens} and the D4-brane charge \eqref{D4dens}.

\section{The ``Abelian'' picture}\label{sec:abel}

In the previous section we constructed a family of eight-supercharge configurations with D4, D2 and D0 charges, from the non-Abelian D$0$ perspective. Following the same logic as in \cite{Bena:2016oqr}, we now want to work out the corresponding  D$4$-brane picture for these configurations. As we explained in the Introduction, we will only construct a macroscopic map between these pictures, by building a D4 configuration that has the same D0, D2 and D4 charges as that of the previous section.

A system of $N_4$  flat D$4$-branes with non-trivial worldvolume flux can carry D$2$ and D$0$ charges \cite{Douglas:1995bn}, given by the electric couplings to $C_3$ and $C_1$
\begin{equation}\label{eq:D4WZ}
  S^{\textrm{D}4}_{\textrm{WZ}} = \mu_4 \int C_5 + \left(\mu_4 \lambda \int \Tr F_2\right) \int C_3 + \left(\mu_4 \frac{\lambda^2}{2} \int \Tr F_2 \w F_2\right) \int C_1\,,
\end{equation}
in the conventions of \cite{Myers:1999ps}. Just as in Section \ref{sec:set}, we prefer to use dimensionless quantities and define $\tilde{F}_2 \equiv \lambda F_2$. From here on we will exclusively use $\tilde{F}_2$ but drop the tilde, and all integrals are now over boxes with sides of (dimensionless) size $K$. In these conventions, the brane numbers are given by
\begin{equation}\label{eq:4branenr}
  N_2^{ij} = \int \Tr \star \! F_2^{ij}\,,\quad N_0 = \frac{1}{2} \int \Tr F_2 \w F_2\,.
\end{equation}
Much like in the D$0$ picture, this system of branes displays an enhancement of supersymmetry if the trace can be split according to
\begin{equation}\label{eq:enhanceD4}
  \begin{split}
    N_0 &= \frac{1}{2} \int \Trs{N_4} \{\star F_2 \w F_2\}\\
     &= \frac{1}{N_4} \sum_{\textrm{D}2\textrm{-pairs}} \left[\left( \int \Trs{N_4} \{\star F_2\}\right) \times \left(\int \Trs{N_4} \{F_2\}\right) \right] \\
     &= \frac{1}{N_4} \sum_{\textrm{D}2\textrm{-pairs}} N_2^2\,,
  \end{split}
\end{equation}
and our interest here is to prevent this enhancement.

The macroscopic map between the D4 and the D0 descriptions preserves the brane numbers according to\footnote{A similar map, as we use here to identify the D$4$ picture of the D$0$-D$2$-D$4$ state, can be found in \cite{KeskiVakkuri:1997ec}, where they use such a map as a technique to find solutions, and also in \cite{Taylor:1996ik,Ganor:1996zk}, in which they perform four T-dualities along a D$0$-D$2$-D$4$ system.}
\begin{equation}\label{eq:024map}
  \begin{split}
    N_2^{ij} = -i\frac{1}{K^2}\Tr [\Phi_i, \Phi_j]\ &\to\ N_2^{(ij)} = \int \Trs{N_4}\{ \star F_2 \}^{ij}\,,\\
    N_{4} N_0 = -\frac{N_0}{2 K^4} \Tr \epsilon^{ijkl} \Phi_i \Phi_j \Phi_k \Phi_l\ &\to\  N_0 N_4 = \frac{N_4}{2} \int \Trs{N_4} \{F_2 \w F_2\}\,.
  \end{split}
\end{equation}
A three-parameter family of D4 configurations with these charges can be obtained using a constant worldvolume flux of the form
\begin{equation}\label{eq:F2ansatz}
  \begin{split}
    F_{12} = F_{34} &= 0\,,\\
    F_{13} = F_{42} &= \frac{\rho_0}{\rho_4} \left(A_{13} \mathbb{I}_{N_4} + \frac{\gamma}{\sqrt{2}} \Xi\right)\,,\\
    F_{14} = F_{23} &= \frac{\rho_0}{\rho_4} \left( A_{14} \mathbb{I}_{N_4} + \frac{\gamma}{\sqrt{2}} \Xi \right)\,.
  \end{split}
\end{equation}
{where $\Xi$ is any traceless $N_4\times N_4$ matrix with $\Tr \Xi^2 = N_4$. It is easy to verify that the above configurations contain the same amounts of D0, D2 and D4 charges as in \eqref{D0dens}, \eqref{D2dens} and \eqref{D4dens} respectively. However, this constant Ansatz is clearly only applicable if $N_4 > 1$.}

From the D$0$ point of view discussed in the previous section, nothing appears to prevent us from considering a solution to the T-brane equations whose scalar profile gives $N_4=1$. From the D$4$ perspective, however, a constant D4 worldvolume-flux solution cannot be chosen, as it would correspond to a 16-supercharge configuration. One is therefore bound to rely on non-constant flux profiles. If the number of D0-branes were finite, it would have been impossible to describe them from the perspective of a single D4-brane with flux.\footnote{This is due to the well-known fact that there are no Abelian instantons on $\mathbb{R}^4$.} Here, however, the number of D0-branes must be infinite, which allows to relax the finite-action requirement when trying to solve the self-duality equation for the D4-brane flux.

Nevertheless we still believe that there exists no description of our system from the D$4$-brane point of view when $N_4 = 1$, and our argument goes as follows. While we are forced to relax the condition of finite action, we still need to demand that the density of D$0$-branes is finite. This means that either the integral determining the D$0$-brane number scales as $K^4$ -- the same as if the integrand were a constant, or equivalently, as the volume of $\mathbb{R}^4$ -- or the expression for the worldvolume flux must contain explicit $K$ dependence. Any explicit $K$ dependence is ruled out since it would, as we will point out later, fail to produce finite and non-vanishing T-brane dynamics in the $K\to\infty$ limit. Hence we conclude that the D$0$-brane number must make the integral scale as $K^4$ to be a solution of interest. This is in turn not possible since a component of the worldvolume gauge potential must satisfy the Laplace equation if its field-strength is to satisfy the Bianchi identity and the self-duality condition. This indicates that this field-strength and its derivatives in Euclidean coordinates, have to obey the ``maximum principle'', which states that these functions cannot have local extrema. This in turn implies that the function cannot be bounded at infinity (and be regular at finite distances at the same time), and hence must have an integral that scales as $K^{>4}$. Although this constitutes no formal proof, this argument is for us convincing enough to believe that the $N_4 = 1$ solution cannot be described from the D$4$-brane point of view. It would be interesting to look into these discrepancies between the D$0$-brane and D$4$-brane pictures further. We hope to provide more insight into this in future work.

\section{Returning to the original frame}\label{sec:return}

In this section we start from the Abelian D$4$-brane perspective of the previous section and perform two T-dualities in order to return to the original T-brane frame. We will reverse the T-duality performed in Section \ref{sec:set} along the directions $x^{3,4}$, and the resulting system will be a set of intersecting D$2$-branes. The latter will be extended along non-compact two-dimensional planes parameterized by the coordinates $x^1$ and $x^2$. Performing the T-duality along the directions of the worldvolume flux of the previous section (Equation \eqref{eq:F2ansatz}) produces the following set of differential equations
\begin{equation}\label{eq:diff}
  \begin{split}
     &\partial_1 X^3 = - \partial_2 X^4 = \frac{\rho_0}{\rho_4} \left(A_{13} \mathbb{I}_{N_4} + \frac{\gamma}{\sqrt{2}} \Xi\right)\,,\\
   &\partial_1 X^4 = \partial_2 X^3 = \frac{\rho_0}{\rho_4} \left(A_{14} \mathbb{I}_{N_4} + \frac{\gamma}{\sqrt{2}} \Xi\right)\,.
  \end{split}
\end{equation}
This system can be easily integrated and {shown to describe} the embedding
\begin{equation}
  \begin{split}
    X^3 &= \frac{\rho_0}{\rho_4} \left(A_{13} \mathbb{I}_{N_4} + \frac{\gamma}{\sqrt{2}} \Xi\right) x^1 + \frac{\rho_0}{\rho_4} \left(A_{14} \mathbb{I}_{N_4} + \frac{\gamma}{\sqrt{2}} \Xi\right) x^2 + \kappa_1\,,\\
    X^4 &= \frac{\rho_0}{\rho_4} \left(A_{14} \mathbb{I}_{N_4} + \frac{\gamma}{\sqrt{2}} \Xi\right) x^1 - \frac{\rho_0}{\rho_4} \left(A_{13} \mathbb{I}_{N_4} + \frac{\gamma}{\sqrt{2}} \Xi\right) x^2 + \kappa_2\,,\\
    X^1 &= x^1 \mathbb{I}_{N_4}\,,\\
    X^2 &= x^2 \mathbb{I}_{N_4}\,.\\
  \end{split}
\end{equation}
It should be noted that these matrix-valued coordinates describe the embedding of $N_4$ D$2$-branes at once. Furthermore, even though this embedding has a matrix structure, the solution is still Abelian, in the sense that any commutator between the coordinates is zero, $[X^i,X^j]=0$, as long as the integration constants $\kappa_{1,2}$ allow it.

By defining complex coordinates $Z=X^1 + i X^2$ and $W=X^4 + i X^3$, we can write the embedding in a holomorphic way\footnote{With some care, $W$ and $Z$ can be compared to the coordinates with the same labels in \cite{Bena:2016oqr}, although our coordinates are matrices.}
\begin{equation}\label{eq:holo}
  W = C Z + \kappa\,,
\end{equation}
where $C$ and $\kappa$ are given by
\begin{equation}\label{eq:holoexpl}
  C = \frac{\rho_0}{\rho_4} \left(A_{14} \mathbb{I}_{N_4} + \frac{\gamma}{\sqrt{2}} \Xi\right) + i \frac{\rho_0}{\rho_4} \left(A_{13} \mathbb{I}_{N_4} + \frac{\gamma}{\sqrt{2}} \Xi\right)\,,\quad \kappa = \kappa_2 +i\kappa_1\,.
\end{equation}

\begin{figure}
  \begin{center}
  \includegraphics{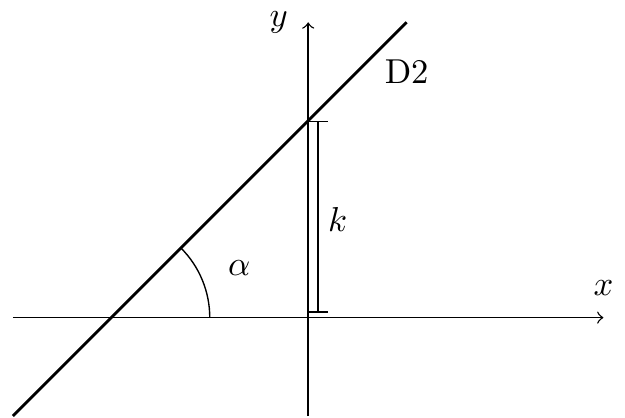}
  \end{center}
  \caption{A projection of Eq.~(\ref{eq:holo}) onto the equation $y = x \tan \alpha + k$, where the angle $\alpha$ would be derived from the constant $C$, and $k$ from $\kappa$.\label{fig:linear}}
\end{figure}

According to Equation (\ref{eq:holo}), the surface over which each D2-brane extends is a flat complex plane embedded in the $\mathbb{C}^2$ parameterized by $Z$ and $W$. Under a certain projection onto a $\mathbb{R}^2$ subspace of $\mathbb{C}^2$, the embedding for one of these branes can be described by Figure \ref{fig:linear}. We see that $A_{ij}$ and $\gamma$ describe the angle the branes make and the integration constants describe shifts of the branes. Note that in the absence of the term proportional to  $\Xi$, the branes would all be parallel and the supersymmetry would be enhanced to 16 supercharges. It is only the parameter $\gamma$ that ensures that the branes are not parallel and hence that the system has only 8 supercharges.

The flat shape of these D2-branes is just a consequence of the constant-flux ``ensemble representative'' solution we chose to focus on in the previous section. A generic member of the ensemble will have non-constant fluxes on the system of D4-branes, which  in turn would give rise to curved D2-brane embeddings after the T-dualities. Hence, the conclusion of our investigation is that the non-Abelian T-brane configuration we started with, made from a stack of D2-branes at $Z=0$ admits an alternative Abelian description consisting in a number of (generally curved) D2-branes intersecting in the $Z,W$ plane.

As we will explain further below, since all the quantities in Equation (\ref{eq:holoexpl}) are independent of $K$ and $N$, our result is finite and hence applies to T-branes with large but finite $N$ as well.

\section{{Discussion}}\label{sec:inter}

Our result confirms the claim made in \cite{Bena:2016oqr} that T-branes admit an alternative Abelian description in terms of branes wrapping holomorphic cycles. We focused on solutions preserving eight supercharges and we restricted to T-branes characterized by constant profiles of the worldvolume scalars, which forced us to consider stacks made of an infinite number of D-branes. The non-commutative scalar profile encodes important physical information, which we extracted and connected to the number of D-branes (of the same dimensionality) needed to describe the system from the Abelian perspective.\footnote{This quantity can be roughly seen as the analogue of the number of Jordan blocks characterizing finite-$N$ T-branes \cite{Bena:2016oqr}.}

The detour we took to link the two pictures allowed us to discover an intriguing connection between the BPS equations governing T-branes and the twenty-year-old Banks-Seiberg-Shenker equation that describe longitudinal five-branes in Matrix Theory. We found a three-parameter family of explicit solutions to these equations and discussed their brane interpretation.

As we have pointed out several times throughout this paper, our matrix-model-inspired construction of T-branes uses infinite matrices, and one may ask whether similar conclusions apply to T-branes made from a finite number of branes. Departing from the infinite-$N$ limit, small $1/N$ corrections are expected to affect our map, and there are at least three sources of such corrections. The first originates from higher-derivative terms in the non-Abelian Born-Infeld action \cite{Constable:1999ac}.\footnote{For D7-branes one could try to extract some of these corrections from known $\alpha'$ corrections in F-theory \cite{Grimm:2013gma,Grimm:2013bha,Minasian:2015bxa}.} The second comes from assuming that the physics of the lower part of Figure \ref{MasterFigure} takes place on an $\mathbb{R}^4$ space. However, for a T-brane in a compactification, one expects this physics to receive corrections of order $1/K$, where $K$ is a typical size of the compactification. When the D$0$-density, $\rho_0$, is finite, $N$ and $K$ are related, and therefore our analysis is precise up to $1/N$ corrections. A third source for $1/N$ corrections (possibly related to the first) is going from the D$0$ to the D$4$ description. The exact relation between the finite- and infinite-$N$ map is still not concrete, and left for future study. However, we believe that the map from the Hitchin system to an Abelian system is valid in general.

In this paper we have limited ourselves to matching the macroscopic charges between the D0-brane and the D4-brane descriptions, and our purpose has not been to find the precise configuration of D4-branes with worldvolume flux providing the alternative Abelian description of the particular D0-brane solution in \eqref{eq:na-sol}. To construct such a microscopic map, one would need to find for example the precise distribution of D0-branes in the non-compact four-dimensional space, encoded by the details of the scalars $\Phi_i$. In analogy with the finite-dimensional systems studied in \cite{Myers:1999ps, Constable:1999ac}, one should be able to reconstruct the ``fuzzy'' distribution from traces of powers of the scalars $\Phi_i$, something which we did not attempt here. For this reason we focused on the easiest possible flux profile reproducing the same macroscopic charges from the D4 perspective, namely the constant-flux solution, which leads to a uniform distribution of D0-branes.  We hope to provide a more refined analysis in a future work.

\section*{Acknowledgements}

We would like to thank Ulf Danielsson, Giuseppe Dibitetto, Mariana Gra\~na, Fernando Marchesano, Washington Taylor, and Angel Uranga for interesting discussions. The work of I.B.~and J.B.~was supported by the John Templeton Foundation Grant 48222 and by the ANR grant Black-dS-String. The work of J.B.~was also supported by the CEA Eurotalents program. The work of R.S.~was supported by the ERC Advanced Grant SPLE under contract ERC-2012-ADG-20120216-320421. In our calculations we have used SymPy \cite{10.7717/peerj-cs.103} and we would like to thank the developers.




\bibliography{refs}

\bibliographystyle{utphysmodb}

\end{document}